\documentclass[12pt]{article}
\usepackage{fullpage}
\usepackage{epsf}
\usepackage{epsfig}
\usepackage{amsthm}
\usepackage{amsfonts}
\usepackage{color}
\usepackage{amsmath}
\usepackage{setspace}
\usepackage{natbib}

\begin{document}

\newcommand{\bm}[1]{\mbox{\boldmath $#1$}}
\newcommand{\mb}[1]{\mathbf{#1}}
\newcommand{\bE}[0]{\mathbb{E}}
\newcommand{\bP}[0]{\mathbb{P}}
\newcommand{\ve}[0]{\varepsilon}
\newcommand{\mN}[0]{\mathcal{N}}
\newcommand{\iidsim}[0]{\stackrel{\mathrm{iid}}{\sim}}
\newcommand{\NA}[0]{{\tt NA}}

\title{\vspace{-1cm} Cases for the nugget in modeling computer experiments}
\author{Robert B.~Gramacy\footnote{Part of this work was done while
    RBG was at the Statistical Laboratory, University of Cambridge}\\
  Booth School of Business\\
  University of Chicago\\
  {\tt rbgramacy@chicagobooth.edu} \and
  Herbert K.~H.~Lee\\
  Applied Math \& Statistics\\
  University of California, Santa Cruz\\
  {\tt herbie@ams.ucsc.edu}}
\date{}

\maketitle

\begin{abstract}
  Most surrogate models for computer experiments are interpolators,
  and the most common interpolator is a Gaussian process (GP)
  that deliberately omits a small-scale (measurement) error term
  called the {\em nugget}.  The explanation is that computer
  experiments are, by definition, ``deterministic'', and so there is
  no measurement error.  We think this is too narrow a focus for a
  computer experiment and a statistically inefficient way to model
  them.  We show that estimating a (non-zero) nugget can lead to
  surrogate models with better statistical properties, such as 
  predictive accuracy and coverage, in a variety of common situations.

\end{abstract}

\section{Introduction}

To some, interpolation is {\em the} defining feature that
distinguishes surrogate models (or emulators) for computer experiments
from models for ordinary experiments.  We think this is old-fashioned
at best and misguided at worst.  It is certainly true that a large
swath of computer experiments are ``deterministic'', in the sense that
once $y(x)$ is known there can be no uncertainty in the output $Y(x')$
if $x' = x$, because the simulator does not behave stochastically.
Interpolation would seem natural in this case, and this is typically
facilitated by a zero nugget in a Gaussian process (GP) prior for
$Y(x)$.  Our first observation is that many of the more recent
computer experiments are indeed stochastic.  A typical formulation is
as an agent based model or finite element simulation where the purpose
is to study cohort/community effects in independent organisms/agents
whose behavior is governed by simple stochastic rules which cannot be
understood analytically.  Examples abound in biology
\citep{johnson:2008,hend:boys:kris:lawl:wilk:2009}, chemistry
\citep{gillespie:2001}, 
and industrial design and engineering
\citep{ankenman:nelson:staum:2009} to name just a few.  It is in this
sense that the defining feature of zero-nugget GPs for computer
experiments is old-fashioned.  Many computer experiments these days
are not deterministic, so in those cases you would include a nugget
without hesitation.  The definition of {\em surrogate model} for a
computer experiment needs to be updated.

But that is not what this paper is really about.  We shall concentrate
on those computer experiments that really are ``deterministic''---in a
sense similar to its usage above but whose decomposition of meaning in
modern experiments is one of the main foci of this paper---and argue
that you should use a nugget anyway.  Our arguments for this are not
computational, although the numerical instabilities of zero-nugget
models are well-documented \citep{abab:bagt:wood:1994,neal:1997}.
Another established criticism of zero-nugget models, upon which we
will not focus, involves theoretical aspects of smoothness and
derivatives.  \citet[pp.~96]{stein:1999} proves that the smoother the
spatial process, the smaller any error or variability needs to be in
order for it to have negligible effect.  Since the standard assumption
in the computer modeling literature is a Gaussian correlation
function, this assumption of infinite differentiability means that the
results are highly sensitive to any possible deviations and thus Stein 
strongly cautions against omitting a nugget term.

As larger nugget values can impact the fitted values of other
parameters \citep{gra:lee:2008b,pepe:2010}, some authors go to great
lengths to reconcile numerical stability and zero-nugget-like
interpolation, usually by using as small a nugget as possible
\citep{ranj:2010}.  Instead, we argue that issues of numerical
stability, while they are strong arguments in favor of a nugget, are a
bit of a red herring in the face of more serious conceptual issues.
We aim to separate the ideology of forcing interpolation from some
important (and undesirable) consequences of the zero-nugget model.  We
shall argue that when the data are sparse or when model assumptions
are violated (e.g., stationarity)---and they typically are---the
nugget is crucial for maintaining good statistical properties for the
emulator (e.g., coverage).  Essentially, when modeling computer
experiments, we must be pragmatic about how assumptions map to
conclusions (surrogate model fits), and this leads us to conclude that
the most sensible default is to estimate a (nonzero) nugget.

The remainder of the paper is outlined as follows.  We conclude this
section with a brief review of GP basics, with further reference to
their application as surrogate models for computer experiments.  In
Section~\ref{sec:break} we elaborate on several conceptual problems
with the zero-nugget approach.  Section~\ref{sec:nug} provides
numerical examples, showing how sparseness of the
sample (Section~\ref{sec:mse}) or violations of standard (and
uncheckable) assumptions (Section~\ref{sec:coverage}) can lead to
inferior predictive surfaces with the zero-nugget approach.  The issue
of ``determinism'' is explored in Section~\ref{sec:det} to similar
effect.  And in Section~\ref{sec:lgbb} we revisit these points on a
real-world computer experiment involving CFD simulations of a rocket
booster re-entering the atmosphere.  Finally, we conclude with a
discussion.

\subsection{GP basics}
\label{sec:gp}

The canonical choice of surrogate model for computer experiments is
the stationary Gaussian process
\citep{sack:welc:mitc:wynn:1989,ohagan99,sant:will:notz:2003}, which
defines a random process whose evaluation at any finite collection of
locations has a multivariate Gaussian distribution with a specified
mean and covariance function that depend only on the relative
positions of the locations.  A typical specification of the covariance
function is the Gaussian correlation, so that the covariance between
any two points is
\begin{equation}
 C(\mb{x}_j, \mb{x}_k) = \sigma^2  K(\mb{x}_j, \mb{x}_k) = 
    \sigma^2 \exp\left\{ - \sum_{\ell=1}^{m}
      \frac{|x_{ij} - x_{ik}|^{2}}{d_\ell}\right\} \,, \label{eq:corr}
\end{equation}
where $m$ is the dimension of the space and $\mb{d}$ is a vector (the
range parameter) which scales the correlation length in each
dimension.  This model will interpolate the data, fitting a smooth
curve between observed outputs of the computer simulator.  The
predictive distribution (or so-called {\em kriging} equations) at new
inputs $\mb{x}^*$ are conditionally Gaussian given $(\mb{x},y)$ pairs
$(\mb{x}_1, y_1), \dots, (\mb{x}_n, y_n)$ and settings of the
parameters $\sigma^2$ and $\mb{d}$.  Our references contain the
relevant equations; we simply remark here that the variance of this
distribution has the distinctive property that it is zero when
$\mb{x}^* = \mb{x}_i$ for one of $i=1, \dots, n$, and increases away
from zero as the distance from $\mb{x}^*$ to the nearest $\mb{x}_i$
increases.  When all elements of $\mb{d}$ are equal, the process is
called {\em isotropic}.

An extension of this model is to include a nugget term,
specifying the covariance function as 
\[
 C(\mb{x}_j, \mb{x}_k) = \sigma^2  K(\mb{x}_j, \mb{x}_k) = 
   \sigma^2 \left[ \exp\left\{ - \sum_{\ell=1}^{m}
     \frac{|x_{ij} - x_{ik}|^{2}}{d_\ell}\right\} + g\delta_{j,k} \right] \,,
\]
where $\delta_{\cdot,\cdot}$ is the Kronecker delta function and $g$
is the nugget term.  Originally introduced to model small-scale
variation in geostatistical models, it is also mathematically
equivalent to adding random noise term into the likelihood.  Thus with
$g>0$, this model no longer interpolates the data, and returns us to a
situation analogous to fitting a mean function with noisy data.  The
predictive distribution has many features in common with that obtained
from the zero-nugget/no-nugget model (above).  For example, the
variance increases away from the nearest input $\mb{x}_i$, but is not
zero at $\mb{x}^* = \mb{x}_i$; rather, it is $g\sigma^2$ at those
locations.

There are many ways to infer the parameters $\sigma^2, \mb{d}, g$
given data $(\mb{x}_1, y_1), \dots, (\mb{x}_n, y_n)$.  For example,
the resulting multivariate normal likelihood can be maximized
numerically, although the presence of the nugget may lead to a bimodal
surface \citep{gra:lee:2008b}.  In this paper we happen to take a
Bayesian approach, but all of our arguments hold true under the
frequentist paradigm as well.  Our implementation is in {\sf R} using
the GP code from the {\tt tgp} library on CRAN \citep{Gramacy:2007}.
The package makes use of a default Inverse--Gamma prior for $\sigma^2$
that yields a multivariate Student-$t$ marginal likelihood
(integrating out $\sigma^2$) for $y_1,\dots, y_n$ given
$\mb{x}_1,\dots,\mb{x}_n$ and the parameters $\mb{d}$ and $g$.
Standard Gamma priors are placed independently on the $d_\ell$
components and $g$, and the Metropolis-within-Gibbs MCMC method is
used to sample from the posterior via the Student-$t$ marginal
likelihood.  The inputs $\mb{x}_i$ are pre-scaled and the proposals
are designed to deal with the bimodal posteriors that can result.

\section{Examining the model assumptions}
\label{sec:break}

Most papers in the literature obsess on the zero-nugget model.  When a
nugget is needed for computational reasons, one aims to make it as
small as possible while still maintaining numerical stability.  The
argument is that the closer the nugget is to zero, the more accurate
the surrogate model approximation is to the computer code output.
This may be true if there is sufficient data, but is it even the right
thing to be worried about?  The measurement error captured by the
nugget (which is presumed to be zero for deterministic computer
simulations) is but one of many possible sources of error.  Here we
discuss four such sources of uncertainty which are likely to be of
greater importance, so it is boggling why so much attention is paid to
the nugget.

\subsubsection*{Simulator bias}

No computer simulator is a perfect representation of the real world.
All simulators are mathematical models and thus only approximate the
real world, so they have some ``bias''.  How we deal with this
discrepancy depends on whether or not real world data are available.
We take those two cases in turn.

When real data are available, it is well-established that the
simulator can be calibrated using the data, that is, the discrepancy
between the simulator and reality can be modeled using an additional
Gaussian process \citep{kennedy:ohagan:2001,sant:will:notz:2003}.
While this addresses the simulator bias, it introduces a source of
noise---that of the real data.  A measurement error term in the
likelihood can be shown to be a re-parameterization of a nugget term
in the covariance function \citep[][Appendix B]{gramacy:2005}.  Thus
we get the same model at the end whether we interpolate then add
noise, or whether we just fit a nugget term.  Since the two approaches
end up at the same place, we might as well embrace the nugget while
fitting the model.

If real data are not available, then the bias cannot be estimated, and
that term is typically ignored and swept under the rug.  Yet pretending
that the simulator is perfect, even though we know it is not is
clearly ignoring a major source of error.  Rather than insist that the
statistical model interpolate the simulator, why not allow the model
to smooth the simulator output?  
We make this point philosophically, in that we do not see why forcing
the model to interpolate something that is deterministically wrong
would be any better than allowing smoothing, and that smoothing can
offer a measure of protection and robustness.

\subsubsection*{The stationarity assumption}

Nearly every analysis in the computer modeling literature makes an
assumption of stationarity, second-order stationarity, or at least
piece-wise stationarity.  More precisely, a {\em residual} process,
arrived at after subtracting off some mean which might be estimated
using a fairly substantial number of regressors
\citep[e.g.,][]{rougier:etal:2009,martin:simpson:2005}, is assumed
stationary.  Such a two-step process can, indeed, lead to good fits.
But it can also be fairly involved as choosing the best regressors in
a non-trivial task, and there is no guarantee that the residuals thus
obtained are stationary.  In any case, the stationarity assumption is
easily challenged.  While it may be reasonable as a close
approximation to the truth, assuming stationarity will not be exactly
correct in most cases.

Typically there is not enough data available to fit a fully
nonstationary model, and if there is enough data, then the model
becomes too difficult to fit efficiently.  Like the bias case, when
there is unknown error, a general statistical principle is that
smoothing (or shrinking) can give better results.  Thus a nugget can
help protect us in the case of moderate deviations from stationarity,
which would be hard to detect in practice.  In
Section~\ref{sec:coverage} we show that even minor violations in the
stationarity assumption lead to emulators with poor statistical
properties without a nugget.

\subsubsection*{Correlation assumptions}

There is an underlying assumption that the specified (typically
Gaussian) correlation structure is correct.  While this is a nice
modeling assumption, it is yet another convenient approximation to reality.
Parameters for the form of the correlation function can be difficult
to fit in practice, and so it is often necessary to simply specify a
reasonable guess.  Since it is only an approximation, this is a
further reason for allowing smoothing in the model.

\subsubsection*{The assumption of a deterministic simulator}

The modeling assumptions addressed above may indeed be reasonable for
a particular true physical process, but the computer implementation of
the solution may still behave in unpredictable ways.  The assumption
of a deterministic simulator may itself be a problem.  Here we discuss
two related possible issues, nonmodelable determinism and theoretical
but not numerical determinism, among other possible problems with the
assumption of deterministic behavior in practice.

Some deterministic functions really are better treated as
nondeterministic.  As a simple example, consider a pseudo-random
number generator where, for any given seed, an output is returned
deterministically (if not also unpredictably unless you know a lot
about numerical analysis).  A version of a computer simulator {\tt
  f$(x)$} approximating a function $g(x)$ numerically might
effectively behave as follows (coded in {\sf R}):

{\singlespacing
\begin{verbatim}
f <- function(x) { 
  set.seed(x)
  return(g(x) + rnorm(1)) 
}
\end{verbatim}
}

\noindent This function {\tt f$(x)$} is theoretically deterministic,
but knowing how it relates to the true function $g(x)$, it would be
irrational to interpolate it.  Clearly one would want to smooth out
the pseudo-random noise which would give us a much better fit the
underlying $g(x)$, and this is exactly what the nugget is designed to
do.  You may argue about the ``deterministic'' nature of the {\tt f}
simulator, but that is to get bogged down in philosophical matters and
miss the practical point.  In this ``cartoon'', $g$ might represent
the mathematical model/equations that describe a system (perhaps only
on paper) whose solutions or realizations aren't available
analytically and require numerically approximate solutions.  This is
where computer implementation in {\tt f} comes in, which may be
deterministic if not otherwise ill-behaved.  In our experience, such
scenarios are more the rule than the exception in modern computer
experiments.

Along similar lines, the {\tt f} and $g$ relationship might just as
easily represent a function with chaotic behavior, which can happen in
complex systems of differential equations, or the Perlin noise
function \citep{perlin:2002}, which is a deterministic method of
generating random-looking smooth surfaces in computer graphics.
Alternatively, the {\tt rnorm(1)} term may stand in for the amount by
which an iterative approximation algorithm steps over the convergence
threshold.  Although we usually we assume that this amount can be made
to be arbitrarily small, this might not always be justified.  One
reason is the lack of uniformity in machine-representable
floating-point numbers.

Now, the above example may seem pathological, but in
Sections~\ref{sec:det} \& \ref{sec:lgbb} we give a synthetic and real
example, respectively, which are essentially the following adaptation:

{\singlespacing 
\begin{verbatim}
f2 <- function(x) { 
  set.seed(x)
  y <- runif(1)
  if(y < 0.9) return(g(x))
  else return(h(x))
}
\end{verbatim}
}

\noindent for some new {\tt h(x)}.  The pseudo-random (but
deterministic) {\tt y} is intended to represent the chance that the
computer code was (poorly) initialized such that it may end up
converging to a sub-optimal (but locally converged) solution {\tt
  h$(x)$} 10\% of the time rather than the true globally converged
approximation to $g(x)$.  This is not an uncommon feature of a modern
computer simulator, i.e., where the final output depends upon an
initial ``solution'' for which there are defaults that usually work,
but sometimes lead to a converged solution which is different from the
one intended.  It is clearly sub-optimal to use a zero-nugget model in
this case, because some of the outputs are not the correct values.
Despite their deterministic nature, we show in Section \ref{sec:det}
that uncertainty about the true function is best modeled with a random
process that smooths rather than interpolates.

\section{Statistically better fits with the nugget}
\label{sec:nug}

\subsection{Protecting against misfits with sparse data}
\label{sec:mse}

Many computer experiments are expensive to run and the number of
datapoints is limited.  As many experiments have higher dimensional
input spaces, the curse of dimensionality implies that the data will
be sparse in the input region.  When the data are sparse,
interpolation can have unpleasant results
\cite[Sec. 2.2]{tadd:lee:gray:grif:2008}.  We present here a simulated
example where the data are sparse in one dimension, but this
represents the concept of sparseness in higher dimensions with a
simpler function or with more data.

\begin{figure}[ht!]
\centering
\begin{minipage}{11.5cm}
\includegraphics[scale=0.45,trim=10 0 0 0]{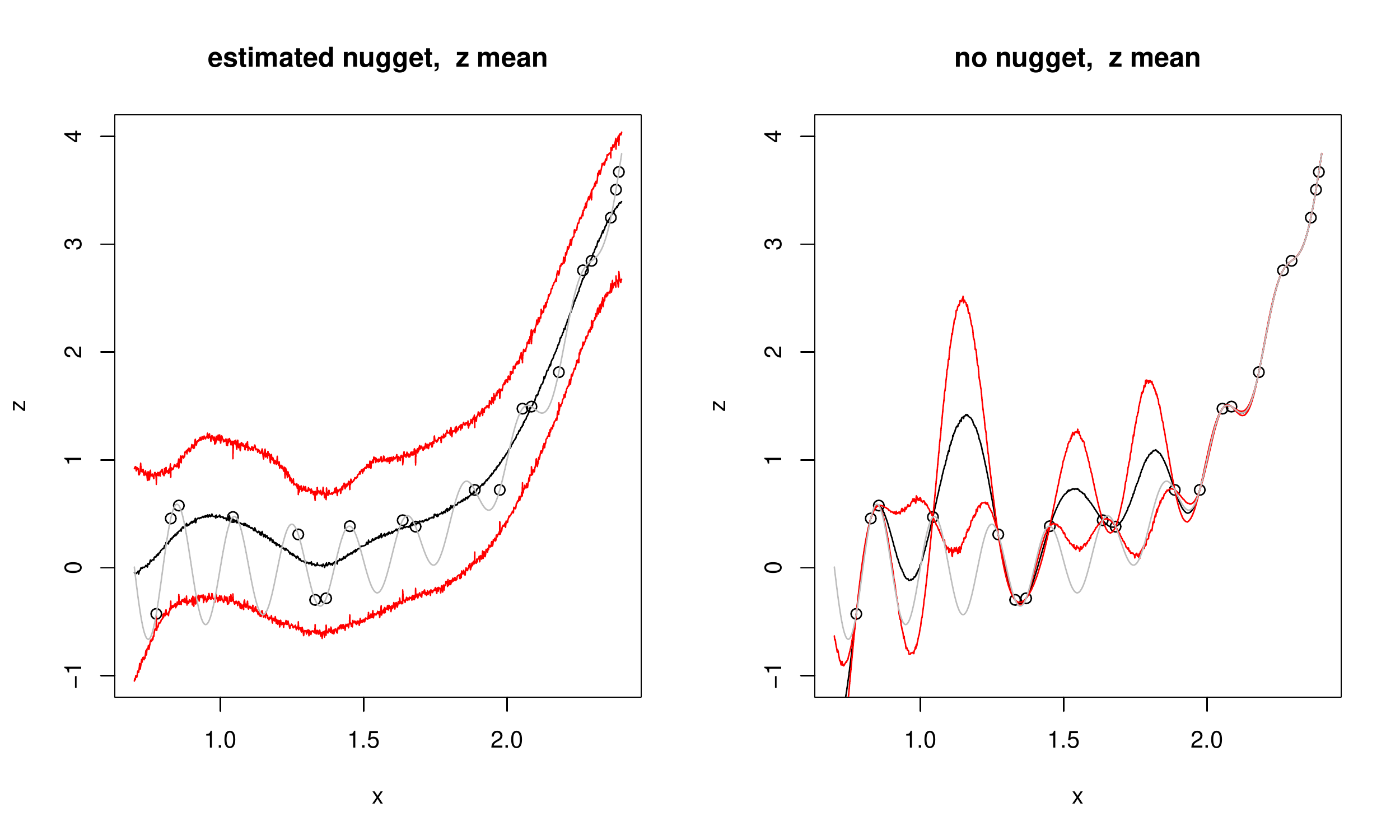}
\end{minipage}
\hfill
\begin{minipage}{4.85cm}
  \begin{tabular}{l|rrr}
MSE &  nug   & nonug \\
\hline
Min.    & 0.0250 & 0.0057 \\
1st Qu. & 0.0999 & 0.0851 \\
Median  & 0.1262 & 0.1399 \\
Mean    & 0.1847 & 0.1929 \\
3rd Qu. & 0.1906 & 0.2290 \\
Max.    & 1.2990 & 1.3510
\end{tabular}
\end{minipage}
\caption{Plots of one example of a fit (dark solid line), confidence 
 bands (red), and true function (light grey). The {\em left} plot shows 
 the fit using a nugget, the {\em right} plot without a nugget.  The 
  {\em table on the right} is the summary of the mean square errors
  under both models for 10,000 repeated uniform designs.}
\label{f:mse}
\end{figure}

Consider the function 
\[
Z = \frac{ \sin(10 \pi X)}{2X} + (X-1)^4 \,.
\] 
Suppose we only have 20 datapoints available (in practice, we would
have more points but more dimensions).  We randomly generated 10000
such datasets (with $X$ generated from a uniform distribution each
time) and fit models both with a nugget and without one.  The table in
Figure~\ref{f:mse} gives the distribution of the mean square errors of
fits under each model, and the model that includes a nugget does
better on average (a paired $t$-test gives a $p$-value of less than
$2.2\times 10^{-16}$).  The plots in Figure~\ref{f:mse} show one of
the runs.  The data are too sparse to get a good fit of the function
for smaller input values.  While the nugget model smooths and produces
reasonable confidence bands, in order to interpolate smoothly the
no-nugget model ends up making predictions well outside the range of
the actual data in that region, and its confidence bands are all over
the place.

In practice, if we only had one-dimensional inputs, we would use
evenly-spaced points for small samples, and much of the issue here
would go away.  However, as the dimension of the space increases, it
becomes impossible to use a regular grid, and random Latin hypercubes
are often used.  It can be impossible to understand how well a design really
covers a high-dimensional space.  Thus the nugget provides good
protection against the strange fits that interpolation can produce.

\subsection{Poor coverage}
\label{sec:coverage}

In fact, the nugget offers protection from a slew of problematic
scenarios.  Here we shall illustrate that the no-nugget model
under-covers the true computer simulator response when the
stationarity assumption is not satisfied.  We use three examples.

\begin{figure}[ht!]
\centering
\begin{minipage}{10.95cm}
\includegraphics[scale=0.45,trim=33 60 0 0, clip=TRUE]{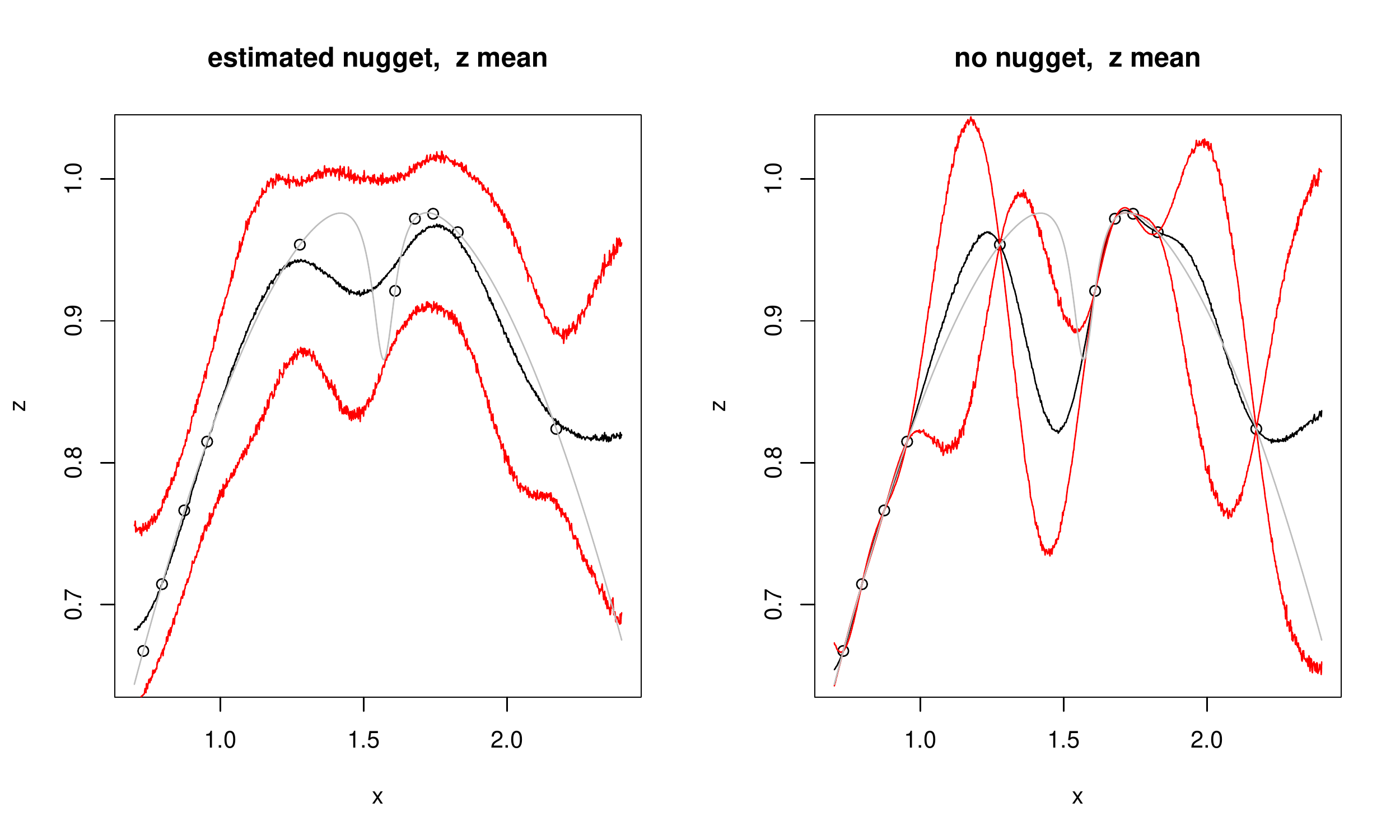}
\vspace{0.2cm}
\includegraphics[scale=0.45,trim=33 0 0 50,clip=TRUE]{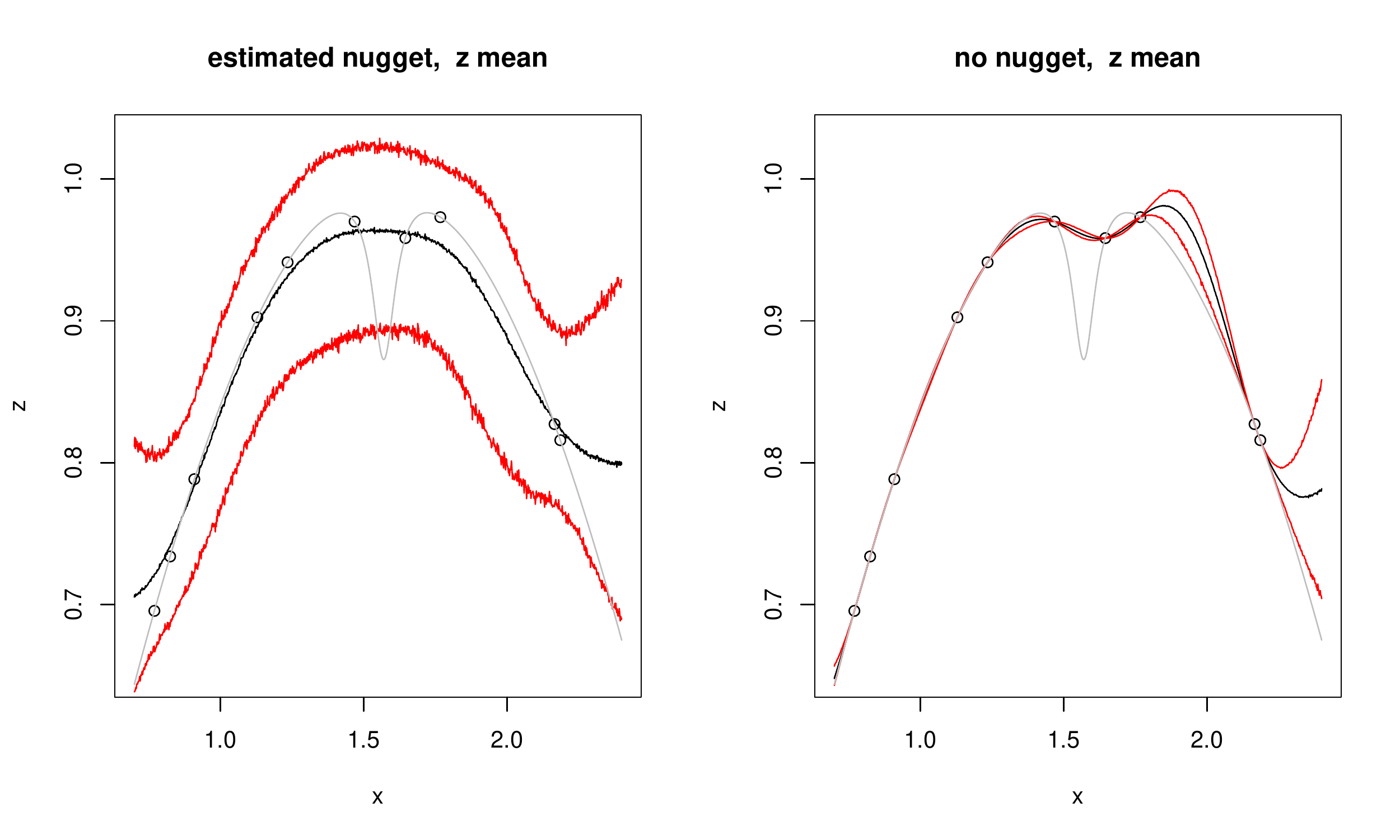}
\end{minipage}
\hfill
\begin{minipage}{5.43cm}
  \begin{tabular}{l|rrr}
coverage&  nug   & nonug \\
\hline
Min.    & 0.3210 & 0.0650 \\
1st Qu. & 0.8028 & 0.5108 \\
Median  & 0.8915 & 0.7230 \\
Mean    & 0.8517 & 0.6531 \\
3rd Qu. & 0.9570 & 0.8068 \\
Max.    & 1.0000 & 0.9960 \\

\\
$\sqrt{\mbox{mah}}$ &  nug   & nonug \\
\hline
Min.    & 7.067 & 20.020 \\
1st Qu. & 16.319 & 47.901 \\
Median  & 25.079 & 91.024 \\
Mean    & 49.928 & 246.772 \\
3rd Qu. & 45.111 & 174.876 \\
Max.    & 560.245 & 2966.92
\end{tabular}
\vspace{0.25cm}

\end{minipage}
\caption{The {\em plots on the left} are examples of fits under under
  two uniform designs in nugget (left column) and no-nugget (right
  column) models. The {\em table on the right} summarizes of the
  coverages and (square root) Mahalanobis distances under both models
  for 100 repeated uniform designs.}
\label{f:coverage}
\end{figure}

The first example is a 1-d function which is clearly nonstationary,
but otherwise mimics typical features of a computer code.  The
response is given by $y(x) = \sin(x) - 0.02 \cdot t_1(x,1.57,0.05)$
where $t_1(\cdot, \mu, \sigma)$ is a Cauchy density with mean $\mu$
and spread $\sigma$.  The two rows of Figure \ref{f:coverage} show
fits for two typical random uniform designs of size ten.  The
difference between smoothing (estimated nugget; {\em left panels}) and
interpolation (no nugget; {\em right panels}) is clear.  We see that
the no-nugget model under-covers the truth (in gray) and can have
wildly different (i.e., narrow or wide) 90\% predictive credible
intervals.  This experiment was repeated 100 times and the percentage
of the area of the input space where the true $y(x)$ was covered
(pointwise) by the 90\% interval was recorded.  A table summarizing
the results numerically is on the right in the figure.  We see from
this table that the under-coverage of the no-nugget model can be
drastic.  For one of the random designs it only covered 6.5\% of
$y(x)$ and $3/4$ of the trials under-covered by more than 10\%.  By
contrast, the model which estimates a nugget has good coverage
properties.  Its median and mean coverages are close to 90\% and the
central 50\% region tightly brackets the truth.  Clearly, connecting
the dots comes at the expense of other, arguably more important,
statistical measures of goodness of fit.

Although it is intuitive, pointwise coverage via might not be an ideal
metric for comparing model fit because it ignores correlation aspects
of the posterior predictive distribution---mis-coverage at one input
is indicative of miscoverage at a continuum of nearby inputs.  A
metric based on the Mahalanobis distance, proposed by
\cite{bastos:ohagan:2009} for GP models, allows such correlation in
the predicted outputs to be taken into account.  For completeness, we
report a summary of the square root of these distances in Figure
\ref{f:coverage} as well.  The conclusions based on this metric are
largely the same: estimating a nugget leads to a superior fit (smaller
distances) compared to the zero-nugget interpolating approach.

\begin{table}[ht!]
\centering
\begin{tabular}{l|rr}
& \multicolumn{2}{c}{exp data} \\
coverage & nug & nonug \\
\hline
Min.    & 0.5479 & 0.3965 \\
1st Qu. & 0.8623 & 0.8242 \\ 
Median  & 0.9185 & 0.8936 \\
Mean    & 0.8962 & 0.8691 \\
3rd Qu. & 0.9492 & 0.9395 \\
Max.    & 1.0000 & 1.0000  \\
\\
$\sqrt{\mbox{mah}}$ & nug & nonug \\
\hline
Min.      & 3.181 & 3.857 \\ 
1st Qu. & 20.870 & 25.850 \\
Median  & 34.840 & 48.270 \\
Mean    & 78.020 &  96.960 \\
3rd Qu. & 67.260 & 100.500 \\
Max.     & 1971.000 & 21820.000 
\end{tabular}
\hspace{1cm}
\begin{tabular}{l|rr}
& \multicolumn{2}{c}{fried data} \\
coverage & nug & nonug \\
\hline
Min.    & 0.5480 & 0.4580 \\ 
1st Qu. & 0.8930 & 0.8350 \\
Median  & 0.9320 & 0.8890 \\
Mean    & 0.9205 & 0.8762 \\
3rd Qu. & 0.9580 & 0.9310 \\
Max.    & 0.9990 & 1.0000 \\
\\
$\sqrt{\mbox{mah}}$ & nug & nonug \\
\hline
Min.    & 14.312 & 15.816 \\ 
1st Qu. & 24.203 & 30.760 \\
Median  & 28.676 & 36.100 \\
Mean    & 29.271 & 36.794 \\
3rd Qu. & 33.621 & 41.968 \\
Max.    & 55.435 & 69.111
\end{tabular}
\caption{{\em Left} are coverages and square-root Mahalanobis distances 
  for the 2-d exponential data; {\em right} for the 5-d Friedman data.}
\label{t:cover}
\end{table}

We performed similar experiments on two higher-dimensional data sets.
The first is a 2-d exponential function $y(x) = x_1 \exp\{-x_1^2 -
x_2^2\}$, which less clearly violates the stationarity assumption.
From the {\em left} side of Table \ref{t:cover} we see a similar
under-coverage of the no-nugget model with repeated uniform designs of
size 20.  Mahalanobis distances are included for completeness. Our
second experiment involved the first Friedman data function
\citep{fried:1991} with five inputs where the response is $y(x) = 10
\sin(\pi x_1 x_2) + 20(x_3 - 0.5)^2 + 10x_4 + 5 x_5$.  This function
is better behaved (i.e., stationarity may be a reasonable assumption).
However, it is apparent that correlation in the response would decay
at different rates along the five coordinates---clear anisotropy.  To
illustrate how the effect of an inappropriate choice of correlation
function is felt more strongly in the no-nugget model we used an
isotropic Gaussian correlation function and uniform designs of size
25.  The results are shown on the {\em right} in the table.

It is worth pointing out that as the size of the designs are
increased, and/or as the data less obviously violate assumptions,
both models (nugget and no-nugget) will tend to 
over cover in practice.  This is because we are fitting a model (GP)
which always yields positive posterior predictive error away from the
(discrete set of) design points, i.e., over an uncountably large
region.  Since the function we are modeling is deterministic, we know
that as the size of the design tends to (countable) infinity we should
be able to obtain a ``perfect'' fit with a high degree polynomial.  So
a GP is the wrong model in this case.  Since over-coverage is
inevitable, under-coverage should be our primary concern, and to avoid
under-covering we can see that a nugget is needed.

\subsection{Challenging determinism in computer simulation}
\label{sec:det}

Some computer experiments are deterministic in a technical sense, but
not necessarily in a way that translates into sensible assumptions for
the building of a surrogate model.  We may reasonably presume that
codes implementing the algorithms and calculations behind the
experiment are nontrivial.  They are expensive to program and
expensive to execute, requiring long iterations to convergence and the
(sometimes arbitrary) specification of tuning parameters, tolerances,
and grid/mesh sizes.  As a rule more than an exception, the resulting
apparatus works better for some choices of inputs than for others.
The most important issue is in detecting global convergence of the
code, whose properties usually depend crucially on other
implementation choices.  It is essentially impossible to guarantee
good global convergence properties, and so this the main target of our
attack on the modeling of such ``deterministic'' computer simulations
without a nugget.

Consider the following computer simulator coded in {\sf R} below.

{\singlespacing 
\begin{verbatim}
## 2-d function
f2d <- function(x1, x2) {
  w <- function(y) {
    return(exp(-(y-1)^2) + exp(-0.8*(y+1)^2) - 0.05*sin(8*(y+0.1)))
  }
  return(-w(x2)*w(x1))
}

## find the minimum of a projection of the 2-d function
f <- function(x) {
  return(optim(par=x, fn=f2d, x2=x)$value)
}
\end{verbatim}}

  \noindent The true underlying function $f(x)$, evaluated by {\tt
    f$(x)$} in {\sf R}, is $\mathrm{arg}\min_{x_1} f(x_1,x)$ where
\begin{align*}
f(x_1,x_2) & = -w(x_1)w(x_2), \;\;\;\;\; \mbox{and} \\
w(y) & = \exp\left(-(y-1)^2\right) + \exp\left(-0.8(y+1)^2\right)
- 0.05\sin\left(8(y+0.1)\right). 
\end{align*}
The optimization method used by the code above is the {\tt optim}
function in {\sf R} initialized at $x_1 = x$.

\begin{figure}[ht!]
\centering
\includegraphics[scale=0.6]{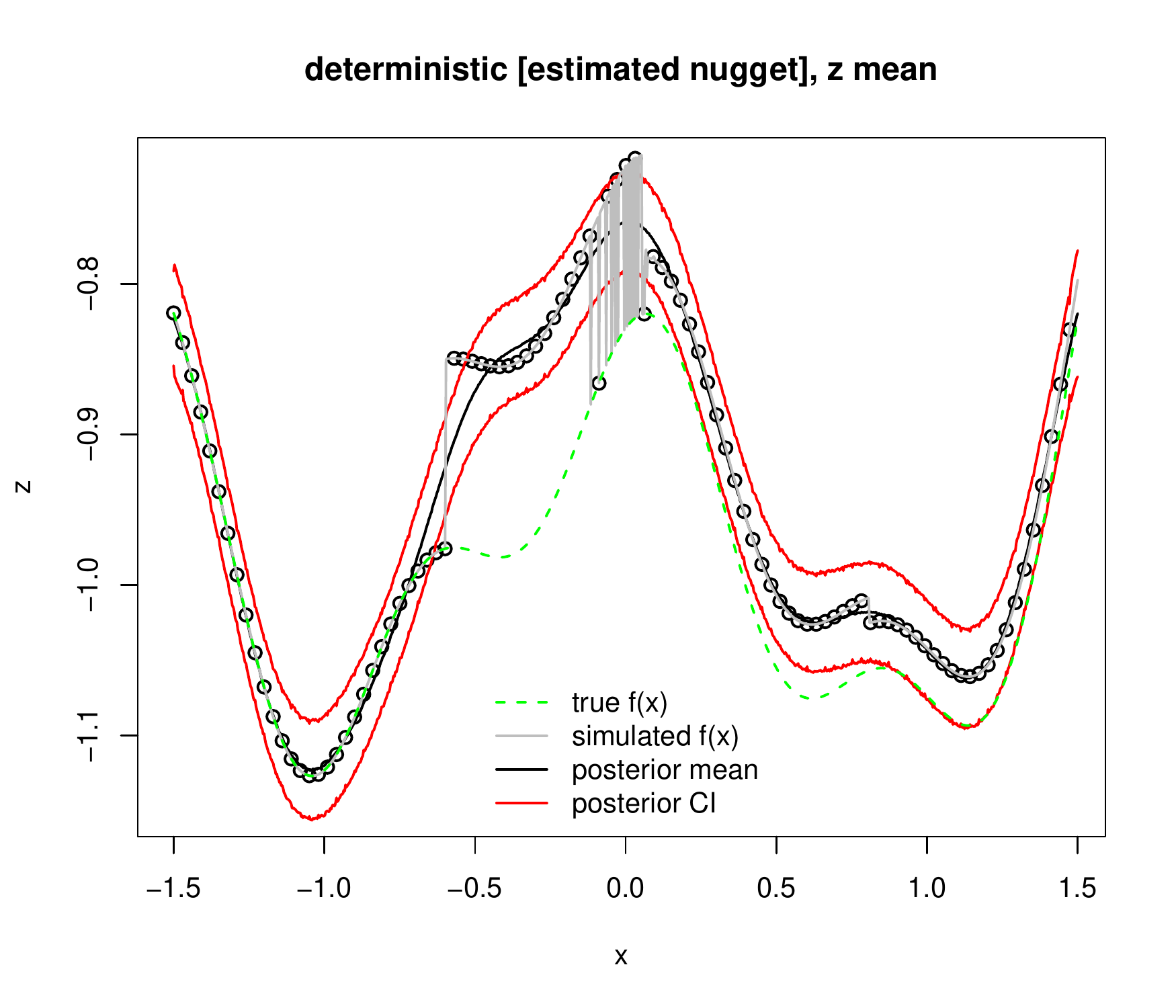}
\caption{GP fit (with an estimated nugget) to a deterministic function
 which is the result of an iterative procedure.}
\label{f:det}
\end{figure}

Figure \ref{f:det} shows the true $f(x)$ (dashed-green) and the output
of the simulator {\tt f$(x)$} (gray) for $x \in [-1.5, 1.5]$.  We can
see that the result of numerically finding the optimal value of the
objective function (initialized somewhat arbitrarily, but not
pathologically) is that the simulations {\tt f$(x)$} are biased, and
behave badly/unpredictably in some parts of the input space. It is
worth noting that both behaviors persist with a different static
initialization scheme; the $f(x_1, x_2)$ surface has about a dozen
local minima.  Also, the implementation {\tt f$(x)$} is completely
deterministic in a technical sense. However, {\tt f$(x)$} is
exhibiting ``random'' behavior of the sort alluded to in Section
\ref{sec:break} as the initialization scheme causes the algorithm to
converge to different local minima in a way that is not (easily)
predictable.  There are three places where the initialization causes
it to have discontinuities (even though the true $f(x)$ is smooth
everywhere), and it is particularly unstable near $x=0$ since $(0,0)$
is more or less equidistant from the many local minima of $f(x_1,
x_2)$ in the 2-d space.

The figure also shows a fit to the computer simulator ({\tt f$(x)$})
output obtained from a gridded design of 100 input--output pairs using
a GP with an estimated nugget.  The fit is sensible given the
discontinuities and otherwise ``noisy'' behavior of the simulator.  It
is not possible to fit this data without a nugget, or even with a
small one, due to numerical instabilities.  However, it is possible to
do so with a reduced design of about 20 points or so.
\begin{table}[ht!]
\vspace{0.25cm}
\centering
\begin{tabular}{l|rr}
  & \multicolumn{2}{c}{computer {\tt f}} \\
  coverage & nug & nonug \\
  \hline
  Min.     & 0.433 & 0.2280 \\
  1st Qu. & 0.787 & 0.6665 \\
  Median  & 0.875 & 0.7345 \\
  Mean    & 0.846 & 0.7276 \\
  3rd Qu. & 0.938 & 0.8362 \\
  Max.    & 0.993 & 0.9760 
\end{tabular}
\hspace{0.75cm}
\begin{tabular}{l|rr|rr}
& \multicolumn{2}{c}{computer {\tt f} } & \multicolumn{2}{|c}{truth
  $f$} \\
$\sqrt{\mbox{mah}}$ & nug & nonug & nug & nonug \\
\hline
Min.     &  14.07 &  36.88  & 41.90 & 279.17 \\
1st Qu. &  33.24 & 139.39 & 104.44 & 8544.02 \\
Median  &  56.87 &  292.17  & 213.78& 12932.55 \\
Mean    &  183.45 &  1163.17  & 855.44 & 15402.05 \\
3rd Qu. &  136.59 &  811.23 & 617.03 & 19392.16 \\
Max.    &  3212.00 & 14561.06 & 12271.41 & 49362.90 \\
\end{tabular}
\caption{Summaries of coverage of the ``deterministic'' computer simulated
  {\tt f(x)} data, and the square-root Mahalanobis distances to {\tt f(x)} and the
  analytic solution $f(x)$.}
\label{t:coverdet}
\end{table}
To connect with the coverage results in the last section (where
stationarity was the issue) we calculated the coverage of {\tt f$(x)$}
with 100 repeated uniform random designs of size 20 under the
estimated nugget and no-nugget models, and the story is much the same
as before.  The results are shown in Table~\ref{t:coverdet}.
Square-root Mahalanobis distances are also shown, as are the distances
to the true $f(x)$.  They show that when ``determinism'' is challenged
as an assumption on the nature of the data-generating mechanism, a
nugget for smoothing is clearly preferred to interpolation in the
surrogate model.

\section{A modern computer experiment}
\label{sec:lgbb}

The Langley Glide-Back Booster (LGBB) is a rocket booster that
underwent design phases at NASA primarily through the use of
computational fluid dynamics simulators that numerically solve the
relevant inviscid Euler equations over a mesh of 1.4 million cells
\citep{rog03}.  The simulator models the forces felt by the rocket at
the moment it is re-entering the atmosphere as a function of three
inputs describing its state: speed (measured by Mach number), angle of
attack (the alpha angle), and sideslip angle (the beta angle).  As a
free body in space, there are six degrees of freedom, so the six
relevant forces/outputs are lift, drag, pitch, side-force, yaw, and
roll.  While theoretically deterministic, the simulator can fail to
converge.  Some nonconvergent runs are caught by an automated checker,
and re-run with a new schedule of initial conditions, but some are
erroneously accepted even after converging to a clearly inferior
solution. Input configurations arbitrarily close to one another can
fail to achieve the same estimated convergence, even after satisfying
the same stopping criterion.

Here we focus on the roll force output on a data set
comprised of simulator runs at 3041 locations.
\begin{figure}[ht!]
\centering
\includegraphics[scale=0.54,trim=60 50 10 40,clip=TRUE]{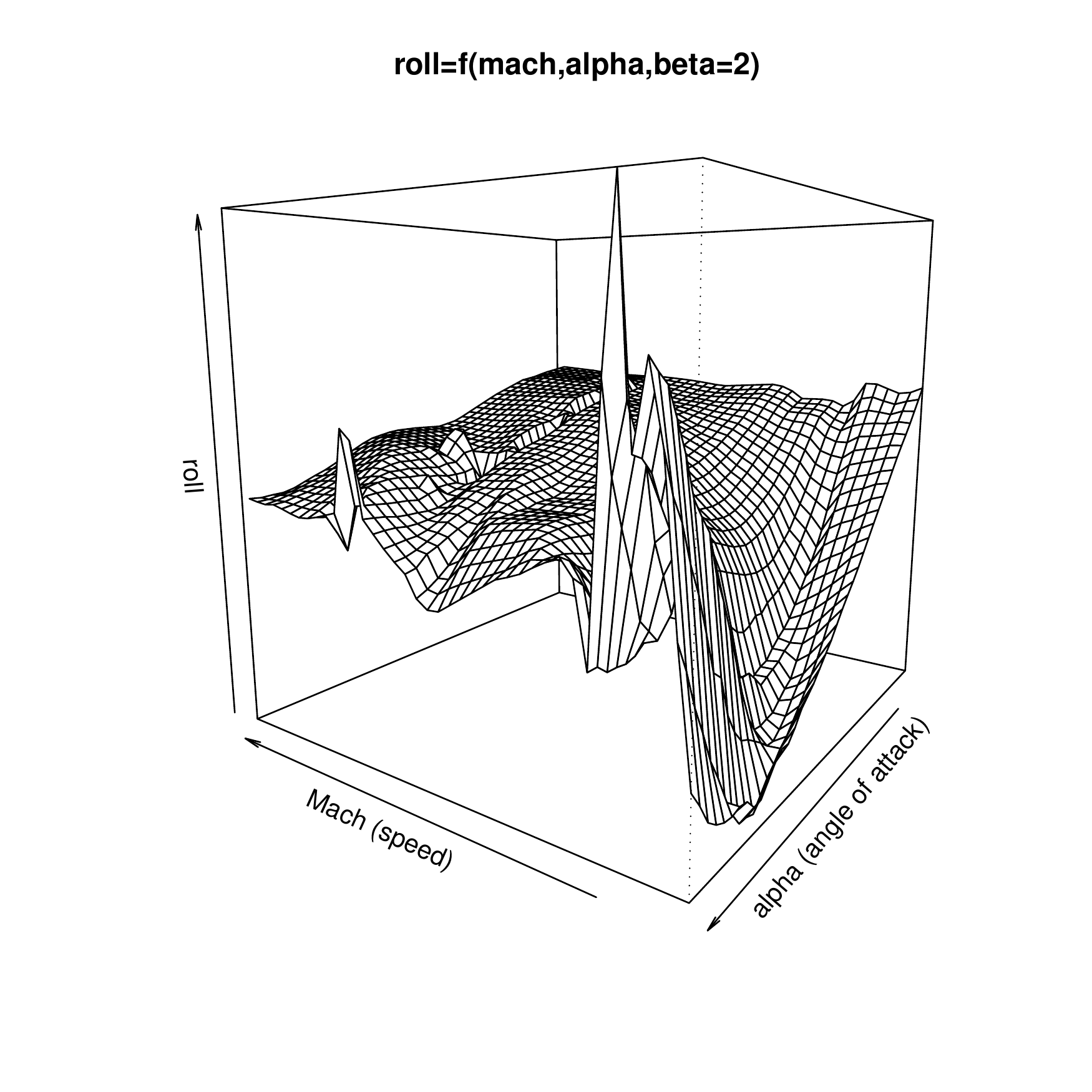}
\includegraphics[scale=0.46,trim=0 0 15 40,clip=TRUE]{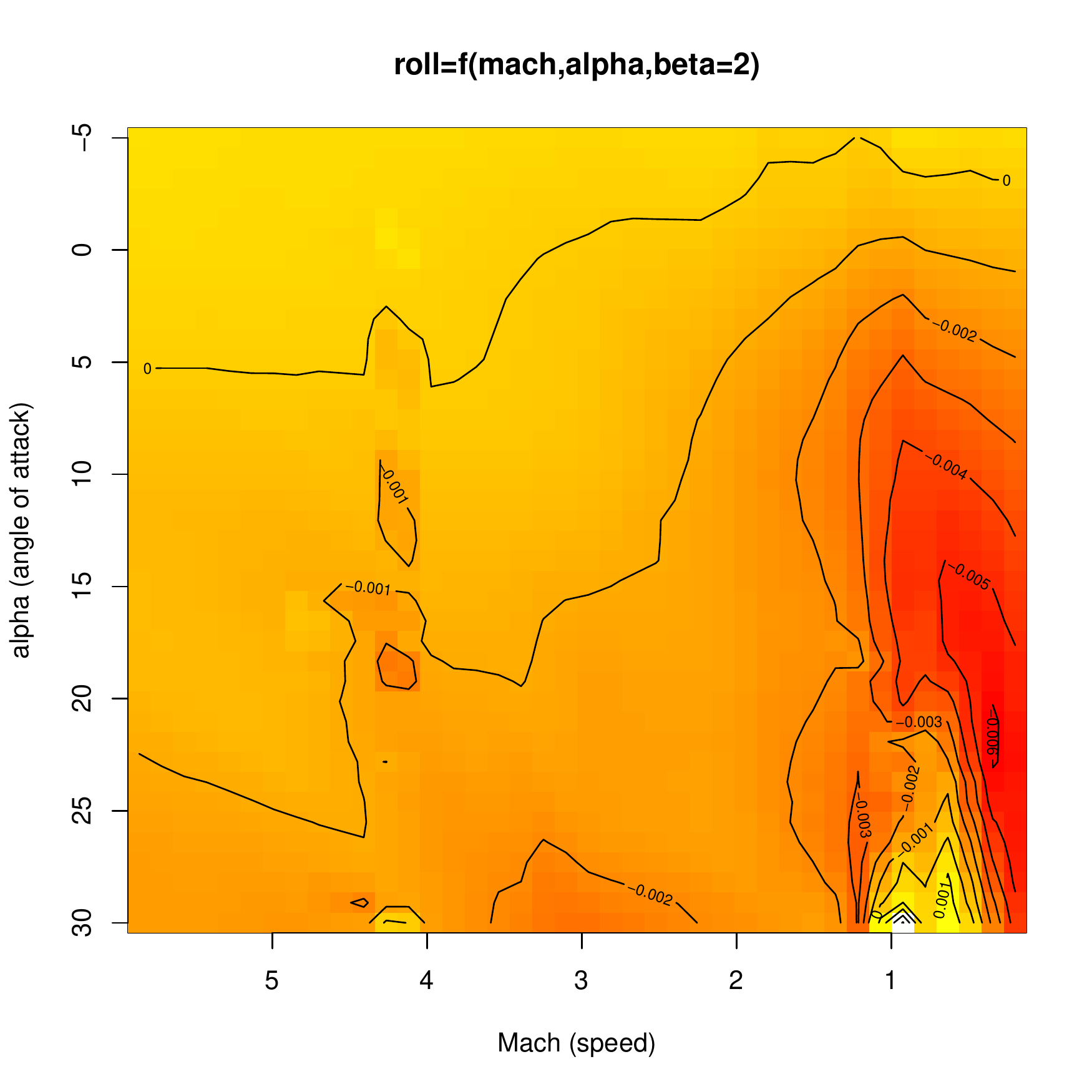}
\caption{Linearly interpolated slice of the roll response plotted in
  perspective {\em (left)} and image/contour {\em (right)} as a
  function of speed (Mach) and angle of attach (alpha), with the slide
  slip angle (beta) fixed to 2.  In the image plot, dark/red values
  are lower and light/yellow values higher in the image plot; the
  perspective plot is rotated for visualization purposes so that the
  closest corner corresponds to low speed and high angle of attack.}
\label{f:roll}
\end{figure}
See Figure~\ref{f:roll} for a 2-d slice of this response.  Previous
work has focused on the lift force \citep{gra:lee:2008} which
exhibited many similar features, and on a sequential design task
taking account of all outputs simultaneously \citep{gra:lee:2009}.
The experimental design is a combination of an initial grid followed
by two hand-designed finer grids focused around Mach one, as the
initial run showed that the most interesting part of the input space
was generally around the sound barrier, where the physics behind the
simulator changes abruptly from a subsonic regime to a supersonic one.
What happens close to and along the boundary is the most difficult
part of the simulation.  The regime changes across this boundary cause
the stationarity assumption to be violated.  Also note the string of
anomalies around Mach four, which appear to converge to local, rather
than global, solutions. So this experiment comprises two challenging
aspects---impractical determinism due to convergence issues and failed
assumptions of stationarity due to physical regime changes---and we
aim to show that the nugget is important in mitigating their effects
when building a surrogate model.

\begin{table}[ht!]
\centering
\begin{tabular}{l|rr}
& \multicolumn{2}{c}{GP} \\
coverage & nug & nonug \\
\hline
Min.     & 0.7547 & 0.5726 \\
1st Qu. & 0.9022 & 0.8427 \\
Median  & 0.9239 & 0.8793 \\
Mean    & 0.9187 & 0.8703 \\
3rd Qu. & 0.9396 & 0.9059 \\
Max.    & 0.9777 & 0.9741 
\end{tabular}
\hspace{1cm}
\begin{tabular}{l|rr}
& \multicolumn{2}{c}{TGP} \\
coverage & nug & nonug \\
\hline
Min.    & 0.7627 & 0.5180 \\ 
1st Qu. & 0.8757 & 0.7195 \\
Median  & 0.8978 & 0.7670 \\
Mean    & 0.8954 & 0.7606 \\
3rd Qu. & 0.9186 & 0.8051 \\
Max.    & 0.9771 & 0.9305
\end{tabular}

\caption{Coverage of the roll response for the LGBB computer
  experiment data using a Gaussian process {\em (left)} and a treed
  Gaussian process {\em (right)}.}
\label{t:rollcov}
\end{table}

Towards this end we calculated the coverages of predictive surfaces
obtained with and without the nugget on a 20-fold partition of the
3041 input/output pairs.  We iterated over the folds, training on
$1/20^{\mathrm{th}}$ of the data, about 159 pairs, and predicting at
the remaining 3009-odd locations in an (inverse) cross-validation
fashion.  The results of this experiment, repeated 100 times for 2000
total coverages for each predictor, are shown on the {\em left} in
Table~\ref{t:rollcov}.  Note that this is not a uniform coverage rate
(over the input area), since the design is more heavily concentrated
around Mach one.  However, the results here are as expected.  The
no-nugget model can severely under-cover in certain examples (with
coverage as low as 57\%), and gives the target coverage of 90\% less
than 1/4 of the time.  The model using an estimated nugget is much
better behaved.  However, it does seem to slightly over-cover.
Although less of a concern, we think that the main cause of this is
the nonuniformity of the design and our choice of priors for {\em
  both} the range and nugget parameters in the face of nonstationarity
and nonconvergence issues.

The treed Gaussian process \citep[TGP,][]{gra:lee:2008} model was
designed to handle the axis-aligned nonstationarity that arises due to
regime changes---exactly the sort exhibited by this data.  In essence,
the TGP model learns an axis-aligned partition of the data wherein the
process is well-fit by separate stationary GP models.  We performed an
identical experiment using TGP and the results are summarized on the
{\em right} in Table \ref{t:rollcov}.  We can see that the coverage of
the version of TGP which estimates the nugget is improved (with better
centering around 90\%), but the no-nugget version is not (showing a
more consistent tendency to undercover).  We are left with the
impression that the nugget is even more important when a
nonstationary model is used, especially in the case of nonconvergent
computer experiments where the assumption of ``determinism'', while
technically valid, may be challenged from a practical standpoint.

\section{Discussion}
\label{sec:discuss}

Several authors have previously argued in favor of a nugget term for
reasons of numerical stability even when fitting a deterministic
model.  We go well beyond numerical convenience, raising fundamental
issues of a variety of modeling assumptions and argue that the use of
a nugget helps protect against poor fits when they are violated. 

In many ways, the themes harped on in this paper are just reminders of
the usual statistical insights, like exploiting bias variance
tradeoffs and shrinkage, which are well-known to lead to improved
estimators and predictors.  It is true that some applications call for
specific features in estimators/predictors, like interpolation for
deterministic computer experiments, which might cause us to eschew
those good practices.  At first glance it would seem like forcing a
zero nugget, versus estimating one, is just one of these situations.
But our experience with fitting GPs to real computer simulation output
has shown us time and time again that the ideal of an interpolating
(zero-nugget) emulator is not ideal at all.  In addition to showing,
in this paper, a subset of examples illustrating why we prefer the
nugget for statistical purposes, we have touched on high level
arguments as to why the zero-nugget model will be inferior in
practice.

In reflecting further on these arguments, thanks to helpful comments
from our referees, we have come to the following perspective on the
matter.  The idealized model for the experiment is the following.
\[
  \mbox{real data from physical process} \stackrel{\mbox{(a)}}{\mbox{
        --- }} \mbox{computer model}
\]
Computer model realizations are expensive so we deploy an emulator [at
(a)] based on GP interpolation of the computer model output to save on
cost/time.  The reality is probably more like the following.
\[
\mbox{real data} \stackrel{\mbox{(b)}}{\mbox{
        --- }} \mbox{physical process} \stackrel{\mbox{(c)}}{\mbox{
        --- }} \mbox{mathematical model} \stackrel{\mbox{(d)}}{\mbox{
        --- }} \mbox{computer code} \stackrel{\mbox{(e)}}{\mbox{
        --- }} \mbox{emulator}
\]
In other words, the real data is a measurement of the physical
process, so there is noise or measurement error [at (b)].  This is not
controversial.  Now, the the computer model is actually two things in
one.  It is a mathematical model which approximates [(c)] the physical
model, and although it does not perfectly describe the system---it is
{\em biased}---it is still an idealization.  The computer
implementation adds a further layer of ``approximation'' [(d)], often with
erratic if still technically deterministic behavior, which can be an
unavoidable nuisance.  We try to find the knobs/settings in the code
that give the best results, but it is never perfect.  
If we interpolate [(e)] the output of the computer code, then we are
interpolating its idiosyncrasies.  If we use a nugget, we might not be
precisely where we want to be (closer to the mathematical model and
thus the physical process [at (c)]), but we have a fighting chance
since we'll smooth out the ``rough edges'' in the computer code.
While this schematic representation is a simplification, it shows the
multiple stages where approximations are made and our empirical work
on real and synthetic data suggest that there is something to it.

Instead of using a nugget, \cite{rougier:etal:2009b} advocate using a
``rougher''---but still interpolating---correlation function like the
Mat\`ern \citep[e.g.,][]{stein:1999}.  This may lead to an improved
fit, and better numerical properties/decompositions of matrices,
compared to smoother correlation functions like the Gaussian
(\ref{eq:corr}).  There are a few complications with this approach,
like the burden specifying an extra smoothing parameter which is hard
to infer statistically.  But more importantly, one wonders whether
better interpolations of idiosyncratic computer code is really what
you want?  We think it would be better to use a nugget and smoother
correlation function because, in our experience, the true solutions to
the mathematical equations underlying the model are well-behaved,
i.e., smooth.  It is also just an easier default option.

\subsection*{Acknowledgments}

We thank Tim Gustafson for introducing us to the Perlin function.
This work was partially supported by National Science Foundation grant
DMS-0906720 and EPSRC grant EP/D065704/1.  We would also like to thank
two referees for thoughtful comments that lead to improvements in the
manuscript.

\bibliography{nugget,../btgpm/tgp}

\begin{thebibliography}{26}
\newcommand{\enquote}[1]{``#1''}
\expandafter\ifx\csname natexlab\endcsname\relax\def\natexlab#1{#1}\fi

\bibitem[\protect\citename{Ababou et~al., }1994]{abab:bagt:wood:1994}
Ababou, R., Bagtzoglou, A.~C., and Wood, E.~F. (1994).
\newblock \enquote{On the Condition Number of Covariance Matrices in Kriging,
  Estimation, and Simulation of Random Fields.}
\newblock {\em Mathematical Geology\/}, 26, 1, 99--133.

\bibitem[\protect\citename{Ankenman et~al., }2009]{ankenman:nelson:staum:2009}
Ankenman, B., Nelson, B., and Staum, J. (2009).
\newblock \enquote{Stochastic kriging for simulation metamodeling.
  Forthcoming.}
\newblock {\em Operations Research\/}.
\newblock To appear.

\bibitem[\protect\citename{Bastos and O'Hagan, }2009]{bastos:ohagan:2009}
Bastos, L. and O'Hagan, A. (2009).
\newblock \enquote{Diagnostics for {G}aussian Process Emulators.}
\newblock {\em Technometrics\/}, 51, 4, 425--438.

\bibitem[\protect\citename{Friedman, }1991]{fried:1991}
Friedman, J.~H. (1991).
\newblock \enquote{Multivariate Adaptive Regression Splines.}
\newblock {\em Annals of Statistics\/}, 19, No. 1, 1--67.

\bibitem[\protect\citename{Gillespie, }2001]{gillespie:2001}
Gillespie, D. (2001).
\newblock \enquote{Approximate accelerated stochastic simulation of chemically
  reacting systems.}
\newblock {\em Journal of Chemical Physics\/}, 115, 4, 1716--1733.

\bibitem[\protect\citename{Gramacy, }2005]{gramacy:2005}
Gramacy, R.~B. (2005).
\newblock \enquote{Bayesian Treed Gaussian Process Models.}
\newblock Ph.D. thesis, University of California, Santa Cruz.

\bibitem[\protect\citename{Gramacy, }2007]{Gramacy:2007}
--- (2007).
\newblock \enquote{{\tt tgp}: An {{\sf R}} Package for {B}ayesian
  Nonstationary, Semiparametric Nonlinear Regression and Design by Treed
  Gaussian Process Models.}
\newblock {\em Journal of Statistical Software\/}, 19, 9.

\bibitem[\protect\citename{Gramacy and Lee, }2008{\natexlab{a}}]{gra:lee:2008}
Gramacy, R.~B. and Lee, H. K.~H. (2008{\natexlab{a}}).
\newblock \enquote{Bayesian treed {G}aussian process models with an application
  to computer modeling.}
\newblock {\em Journal of the American Statistical Association\/}, 103,
  1119--1130.

\bibitem[\protect\citename{Gramacy and Lee, }2008{\natexlab{b}}]{gra:lee:2008b}
--- (2008{\natexlab{b}}).
\newblock \enquote{Gaussian Processes and Limiting Linear Models.}
\newblock {\em Computational Statistics and Data Analysis\/}, 53, 123--136.

\bibitem[\protect\citename{Gramacy and Lee, }2009]{gra:lee:2009}
--- (2009).
\newblock \enquote{Adaptive Design and Analysis of Supercomputer Experiment.}
\newblock {\em Technometrics\/}, 51, 2, 130--145.

\bibitem[\protect\citename{Henderson et~al.,
  }2009]{hend:boys:kris:lawl:wilk:2009}
Henderson, D.~A., Boys, R.~J., Krishnan, K.~J., Lawless, C., and Wilkinson,
  D.~J. (2009).
\newblock \enquote{Bayesian emulation and calibration of a stochastic computer
  model of mitochondrial {DNA} deletions in substantia nigra neurons.}
\newblock {\em Journal of the American Statistical Association\/}, 104, 485,
  76--87.

\bibitem[\protect\citename{Johnson, }2008]{johnson:2008}
Johnson, L. (2008).
\newblock \enquote{Microcolony and Biofilm Formation as a Survival Strategy for
  Bacteria.}
\newblock {\em Journal of Theoretical Biology\/}, 251, 24--34.

\bibitem[\protect\citename{Kennedy and O'Hagan, }2001]{kennedy:ohagan:2001}
Kennedy, M. and O'Hagan, A. (2001).
\newblock \enquote{Bayesian Calibration of Computer Models (with discussion).}
\newblock {\em Journal of the Royal Statistical Society, Series B\/}, 63,
  425--464.

\bibitem[\protect\citename{Martin and Simpson, }2005]{martin:simpson:2005}
Martin, J. and Simpson, T. (2005).
\newblock \enquote{Use of kriging models to approximate deterministic computer
  models.}
\newblock {\em AIAA Journal\/}, 43, 4, 853--863.

\bibitem[\protect\citename{Neal, }1997]{neal:1997}
Neal, R.~M. (1997).
\newblock \enquote{Monte Carlo implementation of Gaussian process models for
  Bayesian regression and classification.}
\newblock Tech. Rep. 9702, Deptartment of Statistics, University of Toronto.

\bibitem[\protect\citename{O'Hagan et~al., }1999]{ohagan99}
O'Hagan, A., Kennedy, M.~C., and Oakley, J.~E. (1999).
\newblock \enquote{Uncertainty Analysis and Other Inference Tools for Complex
  Computer Codes.}
\newblock In {\em Bayesian Statistics 6\/}, eds. J.~M. Bernardo, J.~O. Berger,
  A.~Dawid, and A.~Smith,  503--524. Oxford University Press.

\bibitem[\protect\citename{Pepelyshev, }2010]{pepe:2010}
Pepelyshev, A. (2010).
\newblock \enquote{The Role of the Nugget Term in the Gaussian Process Method.}
\newblock In {\em MODA 9 -- Advances in Model-Oriented Design and Analysis\/},
  149--156. Berlin: Springer-Verlag.

\bibitem[\protect\citename{Perlin, }2002]{perlin:2002}
Perlin, K. (2002).
\newblock \enquote{Improving Noise.}
\newblock {\em ACM Transactions on Graphics\/}, 21, 681--682.

\bibitem[\protect\citename{Ranjan et~al., }2010]{ranj:2010}
Ranjan, P., Haynes, R., and Karsten, R. (2010).
\newblock \enquote{Gaussian Process Models and Interpolators for Deterministic
  Computer Simulators.}
\newblock Department of Mathematics and Statistics, Acadia University.

\bibitem[\protect\citename{Rogers et~al., }2003]{rog03}
Rogers, S.~E., Aftosmis, M.~J., Pandya, S.~A., N.~M.~Chaderjian, E. T.~T., and
  Ahmad, J.~U. (2003).
\newblock \enquote{Automated {CFD} Parameter Studies on Distributed Parallel
  Computers.}
\newblock In {\em 16th AIAA Computational Fluid Dynamics Conference\/}.
\newblock AIAA Paper 2003-4229.

\bibitem[\protect\citename{Rougier et~al.,
  }2009{\natexlab{a}}]{rougier:etal:2009b}
Rougier, J., Guillas, S., Maute, A., and Richmond, A. (2009{\natexlab{a}}).
\newblock \enquote{Expert Knowledge and Multivariate Emulation: The
  Thermosphere--Ionosphere Electrodynamics General Circulation Model
  (TIE-GCM).}
\newblock {\em Technometrics\/}, 51, 4, 414--424.

\bibitem[\protect\citename{Rougier et~al.,
  }2009{\natexlab{b}}]{rougier:etal:2009}
--- (2009{\natexlab{b}}).
\newblock \enquote{Expert Knowledge and Multivariate Emulation: The
  Thermosphere-Ionosphere Electrodynamics General Circulation Model (TIE-GCM).}
\newblock {\em Technometrics\/}, 51, 4, 414--424.

\bibitem[\protect\citename{Sacks et~al., }1989]{sack:welc:mitc:wynn:1989}
Sacks, J., Welch, W.~J., Mitchell, T.~J., and Wynn, H.~P. (1989).
\newblock \enquote{Design and Analysis of Computer Experiments.}
\newblock {\em Statistical Science\/}, 4, 409--435.

\bibitem[\protect\citename{Santner et~al., }2003]{sant:will:notz:2003}
Santner, T.~J., Williams, B.~J., and Notz, W.~I. (2003).
\newblock {\em The Design and Analysis of Computer Experiments\/}.
\newblock New York, NY: Springer-Verlag.

\bibitem[\protect\citename{Stein, }1999]{stein:1999}
Stein, M.~L. (1999).
\newblock {\em Interpolation of Spatial Data\/}.
\newblock New York, NY: Springer.

\bibitem[\protect\citename{Taddy et~al., }2008]{tadd:lee:gray:grif:2008}
Taddy, M., Lee, H. K.~H., Gray, G.~A., and Griffin, J.~D. (2008).
\newblock \enquote{Bayesian Guided Pattern Search for Robust Local
  Optimization.}
\newblock Tech. Rep. ams2008-02, University of California, Santa Cruz,
  Department of Applied Mathematics and Statistics.

\end{thebibliography}
\bibliographystyle{jasa}

\end{document}